\documentclass[acmsmall]{acmart}

\usepackage{listings}
\usepackage[skins]{tcolorbox}
\usepackage{fancybox}
\usepackage{booktabs}

\usepackage{nicematrix}
\usepackage{makecell,hhline}

\usepackage{natbib}

\AtBeginDocument{%
  \providecommand\BibTeX{{%
    \normalfont B\kern-0.5em{\scshape i\kern-0.25em b}\kern-0.8em\TeX}}}

\usepackage{indentfirst}
\usepackage[linesnumbered,ruled,vlined]{algorithm2e}
\usepackage{amsmath,amsfonts}
\usepackage{graphicx}
\usepackage{svg}
\usepackage{textcomp}
\usepackage{xcolor}
\usepackage{amsmath}
\usepackage{indentfirst}
\usepackage{mdframed} 
\usepackage{xcolor,colortbl}
\usepackage{caption}
\usepackage[utf8]{inputenc}

\usepackage{tabularx}

\usepackage{makecell}
\usepackage[export]{adjustbox}
\usepackage{xcolor}
\usepackage{float}
\usepackage{color}
\usepackage{tabularx}
\usepackage{numprint}
\usepackage{paralist}
\usepackage{balance}
\usepackage{amsmath}
\usepackage{tablefootnote}
\usepackage{soul}
\usepackage{xcolor,colortbl}
\usepackage[linesnumbered,ruled,vlined]{algorithm2e}
\usepackage[utf8]{inputenc}
\usepackage{tabularx}
\usepackage{algorithmic}

\usepackage{makecell}
\usepackage[export]{adjustbox}
\usepackage{xcolor}
\usepackage{float}
\usepackage{color}
\usepackage{tabularx}
\usepackage{numprint}
\usepackage{paralist}
\usepackage{balance}
\usepackage{amsmath}
\usepackage{subcaption}
\usepackage{tablefootnote}
\usepackage{soul}

\newcommand{\tabitem}{~~\llap{\textbullet}~~}

\usepackage{etoolbox}
\makeatletter
\@ifpackageloaded{hyperref}{}{\usepackage{hyperref}}
\makeatother

\usepackage{graphicx,multirow}
\usepackage{newfloat,caption,subcaption,listings,mdframed,numprint}
\usepackage{xspace}
\usepackage{soul}

\newcommand\revision[1]{{\color{black} {#1}}}

\definecolor{deepblue}{rgb}{0,.2,0.6}
\definecolor{deepgreen}{rgb}{0,0.5,0}
\definecolor{deepchampagne}{rgb}{0.98, 0.84, 0.65}
\definecolor{mintgreen}{rgb}{0.6, 1.0, 0.6}
\definecolor{vividviolet}{rgb}{0.62, 0.0, 1.0}
\definecolor{mangotango}{rgb}{1.0, 0.51, 0.26}
\definecolor{dkgreen}{rgb}{0,0.5,0}
\definecolor{dkred}{rgb}{0.5,0,0}
\definecolor{dkyellow}{HTML}{c49102}
\definecolor{dkorange}{HTML}{A35A00}
\definecolor{dkbrown}{HTML}{80400B}

\definecolor{gray}{rgb}{0.5,0.5,0.5}

\newcolumntype{a}{>{\columncolor{gray!30}}c}

\usepackage{tikz}

\newlength\myindent
\definecolor{bluekeywords}{rgb}{0.13,0.13,1}
\definecolor{greencomments}{rgb}{0,0.55,0.2}
\definecolor{redstrings}{rgb}{0.9,0,0}
\newcolumntype{b}{X}
\newcolumntype{s}{>{\hsize=.5\hsize}X}

\newcommand{\approach}{{\textsc{AdverIntent-Agent}}\xspace}
\newcommand{\agentReason}{{\textsc{Agent}$_\mathtt{reason}$}\xspace}
\newcommand{\agentT}{{\textsc{Agent}$_\mathtt{test}$}\xspace}
\newcommand{\agentRep }{{\textsc{Agent}$_\mathtt{repair}$}\xspace}

\definecolor{Gray}{gray}{0.92}
\definecolor{LightCyan}{rgb}{0.9,1,1}
\newcolumntype{a}{>{\columncolor{Gray}}c}
\newcolumntype{b}{>{\columncolor{LightCyan}}c}

\newcolumntype{g}{>{\columncolor{Gray}}l}

\usepackage{pifont}

\definecolor{dkgreen}{rgb}{0,0.5,0}
\definecolor{dkred}{rgb}{0.5,0,0}
\definecolor{gray}{rgb}{0.5,0.5,0.5}

\lstdefinestyle{javastyle} {
language=Java,
basicstyle=\ttfamily\bfseries\scriptsize,
  morekeywords={virtualinvoke},
  keywordstyle=\color{blue},
  ndkeywordstyle=\color{red},
  commentstyle=\color{dkred},
  stringstyle=\color{dkgreen},
  numbers=left,
  breaklines=true,
  numberstyle=\ttfamily\scriptsize\color{gray},
  stepnumber=1,
  numbersep=10pt,
  backgroundcolor=\color{white},
  tabsize=4,
  showspaces=false,
  showstringspaces=false,
  xleftmargin=.23in,
  escapeinside={<@}{@>},
}

\newcommand\lt[1]{{\lstinline+#1+}}

\DeclareFloatingEnvironment[fileext=frm,placement={!ht},name=Listing,within=section]{listing}

\usepackage{tikz}

\title{Adversarial Reasoning for Repair Based on Inferred Program Intent}

\author{He Ye}
\affiliation{%
  \department{Department of Computer Science}
  \institution{ University College London}
  \country{UK}}
\email{he.ye@ucl.ac.uk}

\author{Aidan Z.H. Yang}
\affiliation{%
 \department{Software and Societal Systems Department, School of Computer Science}
  \institution{ Carnegie Mellon University}
  \country{USA}}

\author{Chang Hu}
\affiliation{%
  \department{School of Computer Science and Engineering}
  \institution{Macau University of Science and Technology}
  \country{China}}

\author{Yanlin Wang}
\affiliation{%
 \department{School of Software Engineering}
  \institution{Sun Yat-sen University}
  \country{China}}

\author{Tao Zhang}
\affiliation{%
  \department{School of Computer Science and Engineering}
  \institution{Macau University of Science and Technology}
  \country{China}}

\author{Claire Le Goues}
\affiliation{%
 \department{Software and Societal Systems Department, School of Computer Science}
  \institution{Carnegie Mellon University}
  \country{USA}}
\email{clegoues@cs.cmu.edu}

\makeatletter
\newcommand{\printfontsize}{The current font size is: \f@size pt}
\makeatother

\begin{document}

\begin{abstract}
Automated program repair (APR) has shown promising results, particularly with the use of neural networks.
Currently, most APR tools focus on code transformations specified by test suites, rather than reasoning about the program's intent and the high-level bug specification. Without a proper understanding of program intent, these tools tend to generate patches that overfit incomplete test suites and fail to reflect the developer's intentions. However, reasoning about program intent is challenging. 

In our work, we propose an approach called \approach, based on critique and adversarial reasoning. Our approach is novel to shift the focus from generating multiple APR patches to inferring multiple potential program intents. 
\revision{Ideally, we aim to infer intents that are, to some extent, adversarial to each other, maximizing the probability that at least one aligns closely with the developer's original intent.}
\approach is a multi-agent approach consisting of three agents: a reasoning agent, a test agent, and a repair agent. First, the reasoning agent generates adversarial program intents along with the corresponding faulty statements. Next, the test agent produces adversarial test cases that align with each inferred intent, constructing oracles that use the same inputs but have different expected outputs. Finally, the repair agent uses dynamic and precise LLM prompts to generate patches that satisfy both the inferred program intent and the generated tests. 

\approach was evaluated on two benchmarks: Defects4J 2.0 and HumanEval-Java. \approach correctly repaired  77 and 105 bugs in both benchmarks, respectively. 
Our work helps reduce the effort required to review patches by enabling developers to assess program intent in natural language, rather than reviewing code patches.

\end{abstract}

\maketitle

\section{Introduction}
\label{sec:introduction}

Automated program repair (APR) aims to reduce the manual and costly processes involved in software maintenance tasks related to bug fixing~\cite{Monperrus2015,TSE-repair-survey}. A wide range of approaches have been proposed to modify programs at the source level to bring behavior in line with a given test suite. These techniques include search-based~\cite{LeGoues2012GenProg,ssFix, sofix,Yuan2017ARJAAR,astor,ali-issta19-bytecode,sharpFix,relifix-shinwei-icse15,APR-Fuzzing-ISSTA23}, semantics-based~\cite{Angelixicse16,acs,JFix, concolic-repair-PLDI21,patch-transplantation-TOSEM21,CrashProgramRepair-ISSTA19}, and learning-based~\cite{codit-tse20,SEQUENCER,RewardRepair-icse22,Tufano-ICSE19,modit-ase21,CoCoNuT,Recoder,ye2022selfapr,ITER} techniques.
Not unsurprisingly, the emergence of Large Language Models (LLMs) has motivated growing interest in using program repair techniques, such as code generated by GitHub Copilot \cite{CopilotingCopilots}, Amazon  Code Whisperer \cite{yeticstiren2023evaluating}, and Replit \cite{fan-automatedicse23}.
Recent large-scale industrial deployments from Google \cite{googlerepair} and Meta \cite{sapfix} 
provide promising initial validation that techniques developed in the lab can fruitfully move to industrial practice.  

However, existing deployments (and research prototypes) remain limited in the types of bugs they successfully target. Meta's SapFix, for example, specifically targets certain types of null pointer exceptions.  This allows tool designers to have more confidence that narrowly tailored patches are likely to be correct and acceptable.  
Acceptability in program repair has long been informed by program behavior on test suites, as tests are common in practice, and well-understood.  
However, overfitting to provided tests~\cite{cure-worse-15} is a known problem, resulting in patches that cause tests to pass, but do not generalize to the true specification, or developer intent.  
Researchers have proposed to tackle overfitting in traditional program repair through post-assessment approaches, such as additional heuristics~\cite{ICSE18-patchsim}, probability models~\cite{ye2021ods,TianASE20,Jooyong-Poracle-TOMSE23}, or test augmentation techniques~\cite{drr,gao2019crash}. 
However, the problem is intrinsic in patch generation, and existing techniques remain limited in the complexity of defects they can empirically correctly tackle. 

We argue that such traditional techniques are limited in particular by their focus on transformation types in their design (such as new templates, or new types of synthesis, to use to attempt various types of repair). They do not, by and large, attempt to reason about higher level intent of a given program. 
\emph{An intent in this work is defined as the expected behavior of a function specification that developers aim to achieve.}
LLM-driven techniques have recently offered an alternative perspective, potentially increasing APR expressiveness and efficiency by both (a) engaging an LLM to reason about a program's intended behavior, bringing developer intent more explicitly into the process, and (b) iteratively refining initial patch attempts towards more complex patches. This line of LLM-based work enables a self-repair loop to refine previously generated intermediate patches, which is considered state-of-the-art in the field~\cite{hidvégi2024cigar,ChatRepair-ISSTA24,ITER}.

This idea is promising, but incomplete: iterative, conversational repair approaches like  ChatRepair~\cite{ChatRepair-ISSTA24}, ITER~\cite{ITER}, and Cigar~\cite{hidvégi2024cigar} fundamentally assume that an initial repair attempt proposed by an LLM can be refined to an acceptable solution. However, LLMs often face challenges in reasoning, and the initial bug-fixing intent is not always correct~\cite{llm-cannot-self-coorrect}. In such cases, increasing the number of conversational iterations is unlikely to help~\cite{is-selfrepair-a-silver-bullet}.
Indeed, during such iterative processes, these approaches often cause conversational agents to issue refusal responses~\cite{wang-acl-2023-know, xie2024sorrybenchsystematicallyevaluatinglarge}. 




In this work, we propose an approach, \approach, consisting of three agents: one for reasoning about program intent, another for generating tests based on that intent, and a third for producing patches that satisfy the inferred program intent. The core idea behind \approach is to adversarially reason through and construct evidence about the developer intent surrounding the buggy code, and use this reasoning to inform patch construction. At a high level, this means that \approach uses LLM-based agents to (a) infer the developer intention and localize a defect, (b) generate tests to validate both those inferred intentions, and produced patches, supplementing developer-provided tests, and (c) iterates on generated patches conversationally to construct high-quality repairs.  More specifically, \approach:

Uses adversarial reasoning and testing to construct and identify the most likely correct intents.
\approach uses LLMs to infer multiple possible developer intentions with respect to a buggy function.  These  program intents are, by construction, intended to be \emph{independent} and \emph{adversarial}: that is, ideally, it should be impossible for the function, when repaired, to satisfy more than one.  
\approach uses test generation to help demonstrate and validate that the inferred intentions are likely adversarial, assuming the intention is correct by construction and thereby sidestepping the oracle problem~\cite{oracleproblem}.
These tests allow for more effective automated reasoning about which intention is correct, and reduces the risk of overfitting to the developer-provided tests.  
The approach also increases the diversity of the considered patch pool by construction, increasing the likelihood of repair success.  
In practice, it is difficult to construct completely adversarial intents. Our work is an early exploration that uses test cases to quantify the degree of adversariality and to maximize the distinctiveness of intents within a limited number of attempts.

\approach uses adversarial reasoning to explore multiple possible program intentions.  However, multiple root causes are still possible, even given a single possible program intent. 
\approach uses dynamic precise prompts to first ask LLMs to reason about the top-k root causes of bugs based on a given inferred intent. We use those root causes to seed additional prompts to request patches that address each root cause.  
This approach contrasts with previous uses of generic prompts for patch refinement \cite{ChatRepair-ISSTA24}, as it explicitly guides the generation of diverse solutions.

We implement \approach and evaluate it against two benchmarks: Defects4J 2.0 \cite{Defects4J} and HumanEval-Java \cite{jiang2023knod}. On the Defects4J 2.0 benchmark, \approach successfully repairs  77 bugs, outperforming  related works \cite{ChatRepair-ISSTA24, alpha-repair-fse22, repairagent, ye2022selfapr}. In HumanEval-Java, \approach achieves 105 bug repairs, also surpassing the considered related works. These findings highlight the effectiveness of \approach in addressing the challenges of automated program repair.

We make the following contributions: 

\begin{itemize}
\item We devise \approach, an adversarial reasoning agent for program intents, to guide the generation of diverse and high-quality patches for program repair.
\item \approach is extensively evaluated on two datasets, including Defects4J v2.0 and HumanEval, achieving state-of-the-art performance.
\item Our work explores an original technique of adversarial program intents, where the degree of adversarial is measured by the number of adversarial test cases that can be generated.  
\end{itemize}

\section{Illustration}
\label{sec:motivation}

   \begin{figure}
\refstepcounter{listing}

\label{lst:illustrative}

\noindent\begin{minipage}[b]{.55\textwidth}
    \begin{lstlisting}
public int countUpper(String s) {
 for(int i=0; i<s.length(); i+=2){
  char c = s.charAt(i);
  if(c=='A'||c=='e'||c=='I'||c=='o'||c=='u') {
     count += 1; 
  }
 return count; 

    \end{lstlisting}
    \subcaption{Buggy program from HumanEval-Java}
    \label{illus-buggy-program}            
    \end{minipage}%
    \hfill
\begin{minipage}[b]{.45\textwidth}
    \begin{lstlisting}
//Test case to expose the bug
public void testCountUpper()  {
 int result = countUpper("aBCdEf");
 assertEquals(1, result);
}
<@\color{dkred}{AssertionError: }@>
<@\color{dkred}{Expected: 1 Actual: 0}@>

    \end{lstlisting} 
    \subcaption{Initial failing test case to expose the bug}
    \label{illus-init-fail-test}            
\end{minipage}
\hfill
\hfill

\begin{minipage}[b]{.99\textwidth}
\begin{minipage}[b]{.33\textwidth}

\begin{tcolorbox}
[colback=white,colframe=dkbrown ,arc=0pt,boxrule=0.5pt,title=Initial Program Intent 1:,boxsep=1pt,left=0pt,right=0pt,top=1pt,bottom=1pt,fonttitle=\small ]
\footnotesize
The program is to count \textbf{every uppercase vowels}  in  `s`. The faulty code is: 

\colorbox{red!40!}{- for (int i=0; i< s.length(); i += 2 \{ }
\colorbox{red!40!}{- if (c == 'A'||c == 'e'||c == 'I'||c == 'o'...}

\hfill
    \label{lst:add-element-patch} 
\end{tcolorbox}           
\end{minipage}
\begin{minipage}[b]{.325\textwidth}
\begin{tcolorbox}
[colback=white,colframe=black,arc=0pt,boxrule=0.5pt,title=Adversarial Program Intent 2:,boxsep=1pt,left=0pt,right=0pt,top=1pt,bottom=1pt,fonttitle=\small ]
\footnotesize
The program is to count uppercase vowels \textbf{at even indices} in string `s`. The faulty code is: 

\colorbox{red!40!}{- if (c == 'A'||c =='e'||c =='I'||c == 'o'... }
    \label{lst:add-element-patch} 
\end{tcolorbox}           
\end{minipage}
\begin{minipage}[b]{.33\textwidth}
\begin{tcolorbox}
[colback=white,colframe=dkred,arc=0pt,boxrule=0.5pt,title=Adversarial Program Intent 3,boxsep=1pt,left=0pt,right=0pt,top=1pt,bottom=1pt,fonttitle=\small ]
\footnotesize
The program is to count \textbf{uppercase and lowercase} vowels at even indices in string `s`.
\colorbox{red!40!}{ - if (c == 'A'||c == 'e'||c == 'I'||c == 'o'...}
    \label{lst:add-element-patch} 
\end{tcolorbox}

\end{minipage}
\subcaption{\agentReason reasons three program intents. }
\hfill

\label{lst:intent-reasoning}  
\end{minipage}

\begin{minipage}[b]{.99\textwidth}
\begin{minipage}[b]{.33\textwidth}
\begin{tcolorbox}
[colback=white,colframe=dkbrown,arc=0pt,boxrule=0.5pt,title= Initial Test Cases 1 :,boxsep=1pt,left=0pt,right=0pt,top=1pt,bottom=1pt,fonttitle=\small ]
\scriptsize
Input: s = \textcolor{black}{"UNIvERsiTy"}, Expected output: \textcolor{blue}{3}

Input: s = \textcolor{black}{"Apple"}, Expected output: \textcolor{blue}{1}

Input: s = \textcolor{black}{"bAnaNa"}, Expected output: \textcolor{blue}{1}

Input: s = \textcolor{black}{"AeioOU"}, Expected output: \textcolor{blue}{3}

\end{tcolorbox}           
\end{minipage}
\begin{minipage}[b]{.33\textwidth}
\begin{tcolorbox}
[colback=white,colframe=black,arc=0pt,boxrule=0.5pt,title= Adversarial Test Cases 2 :,boxsep=1pt,left=0pt,right=0pt,top=1pt,bottom=1pt,fonttitle=\small ]
\scriptsize
Input: s = \textcolor{black}{"UNIvERsiTy"}, Expected output: \textcolor{blue}{3}

Input: s = \textcolor{black}{"Apple"}, Expected output: \textcolor{blue}{1}

Input: s = \textcolor{black}{"bAnaNa"}, Expected output: \textcolor{blue}{0}

Input: s = \textcolor{black}{"AeioOU"}, Expected output: \textcolor{blue}{2}

\end{tcolorbox}           
\end{minipage}
\begin{minipage}[b]{.33\textwidth}
\begin{tcolorbox}
[colback=white,colframe=dkred,arc=0pt,boxrule=0.5pt,title= Adversarial Test Cases 3:,boxsep=1pt,left=0pt,right=0pt,top=1pt,bottom=1pt,fonttitle=\small ]
\scriptsize
Input: s = \textcolor{black}{"UNIvERsiTy"}, Expected output: \textcolor{blue}{3}

Input: s = \textcolor{black}{"Apple"}, Expected output: \textcolor{blue}{2}

Input: s = \textcolor{black}{"bAnaNa"}, Expected output: \textcolor{blue}{0}

Input: s = \textcolor{black}{ "AeioOU"}, Expected output: \textcolor{blue}{4}

\end{tcolorbox}

\end{minipage}

\subcaption{\agentT generates three sets of adversarial tests to measure the degree of adversarial in different program intents based on different outputs over all generated tests: the adversarial score between the first and second intent is 50\%, the adversarial score between first and third intent is 75\%.}
    \label{lst:tests-reasoning}     
\end{minipage}
    
\hfill
\hfill

\begin{minipage}[b]{.99\textwidth}

\begin{minipage}[b]{.33\textwidth}
\begin{tcolorbox}
[colback=white,colframe=dkbrown,arc=0pt,boxrule=0.5pt,title= \scriptsize Patch for  Intent 1 - Overfitting:,boxsep=1pt,left=0pt,right=0pt,top=1pt,bottom=1pt,fonttitle=\footnotesize ]
\scriptsize

\tiny \colorbox{red!40!}{- for (int i=0; i< s.length(); i += 2 \{ }
\tiny \colorbox{green!40!}{+ for (int i=0; i< s.length(); i += 1 \{ }
\tiny \colorbox{red!40!}{- if (c=='A'||c=='e'||c =='I'||c=='o'||c=='u') \{}
\tiny \colorbox{green!40!}{+ if (c=='A'||c=='E'||c =='I'||c=='O'||c=='U') \{}

\end{tcolorbox}           
\end{minipage}
\begin{minipage}[b]{.325\textwidth}
\begin{tcolorbox}
[colback=white,colframe=black,arc=0pt,boxrule=0.5pt,title= Patch for  Intent 2 - Correct:,boxsep=1pt,left=0pt,right=0pt,top=1pt,bottom=1pt,fonttitle=\small ]
\scriptsize
\colorbox{red!40!}{- if (c=='A'||c=='e'||c=='I'||c=='o'||c=='u') \{}
\colorbox{green!40!}{+ if (c=='A'||c=='E'||c=='I'||c=='O'||c=='U') \{}

\hfill
\hfill

\hfill
\hfill

\end{tcolorbox}           
\end{minipage}
\begin{minipage}[b]{.33\textwidth}
\begin{tcolorbox}
[colback=white,colframe=dkred,arc=0pt,boxrule=0.5pt,title=  Patch for  Intent 3 - non-plausible:,boxsep=1pt,left=0pt,right=0pt,top=1pt,bottom=1pt,fonttitle=\footnotesize ]
\scriptsize
\colorbox{red!40!}{- if (c=='A'||c=='e'||c == 'I'||c=='o'||c=='u') \{}
 \colorbox{green!40!}{+ if (c=='A'||c=='E'||c == 'I'||c=='O'||c=='U'}
\tiny \colorbox{green!40!}{ || c=='a'||c=='e'||c == 'i'||c=='o'|| c=='u') \{}

\hfill

\end{tcolorbox}   

\end{minipage}

\subcaption{\agentRep generates three sets of patches to satisfy program intents and pass adversarial tests}
\hfill
\label{lst:repair-patches}  

\end{minipage}

\caption{An illustrative example \approach to show the difference between initial and adversarial reasoning and corresponding patches and test case generated.  }

\end{figure}

In this section, we illustrate the workflow of \approach with an example.
Consider \autoref{illus-buggy-program}, showing a bug in the \texttt{CountUpper} program from the  Humaneval-Java \cite{jiang2023knod} dataset.  We use this example to demonstrate \approach's
initial and adversarial approach to formulate program intention as part of a repair process. 

Given the buggy program   and a test failure that exposes the bug, as shown in  \autoref{illus-init-fail-test}, \approach works as follows. First, a \emph{reasoning} agent (\agentReason) starts with an initial prompt to reason about the intent of program, and its faulty statements: \textit{Summarize the program intent, and identify faulty locations}. These responses  are shown in  \autoref{lst:intent-reasoning}. This produces an \emph{initial} inferred intent and potentially buggy lines (first box of \autoref{lst:intent-reasoning}), which pinpoints the \lt{for} loop and the \lt{if} condition are both incorrect, as it only checks every other letter with mixed lowercase and upper case vowels.

\agentReason also constructs an \emph{adversarial} prompt. This allows \approach to simultaneously explore a second possible intent. It therefore asks the LLM, 
``If the previous intent is NOT correct,...'', producing the \emph{adversarial} intent to count uppercase vowels at every other position (shown in the second box of \autoref{lst:intent-reasoning}).  This second prompt identifies only the \lt{if} condition on line 4 as problematic.  \approach continues this process to achieve another adversarial program intent to count both uppercase and lowercase vowels at even indices (shown in the third box of \autoref{lst:intent-reasoning}). 
Note that this differentiates our approach from prior agent-based work~\cite{repairagent,jiang2023knod} by considering different possibilities for program intents.

Next, \autoref{lst:tests-reasoning} illustrates how the \emph{testing} agent, \agentT, uses the two adversarial intents to generate new differential test cases that reflect them. The goal of \agentT is twofold. First, we use tests to measure the degree of adversarial behavior in the inferred program intents. The prompt instructs the LLM to construct tests with the \emph{same inputs} but expected different outputs for each program intent. A greater number of differentially generated tests based on the same inputs indicates a higher degree of adversarial behavior in the program intents. When the number of tests falls below a predefined threshold, we reject and re-generate the inferred program intent to ensure all intents are adversarial to each other.
In our example in \autoref{lst:tests-reasoning}, the adversarial score between the first and second intents is 50\%, as two tests produce different results out of four total tests. In contrast, the adversarial score between intent 1 and intent 3 is 75\%. Under our settings, all three intents are considered adversarial, as the threshold is set to 33.3\% (the metric is introduced in 
\autoref{sec-app-test}. 
Otherwise, the intents will be re-generated until they achieve the expected degree of adversarialness.

The second goal of \agentT is to generate tests that reflect each intent and specify the correct behavior for patches. Since we assume each individual intent is correct, there is theoretically a solution to an oracle problem. However, practically, ensuring assertion correctness can be challenging. We apply criticism and ranking strategies to alleviate this issue, which are discussed in detail in \autoref{sec-app-test} approach. To the best of our knowledge, \approach is the first technique to generate test cases at a general level during the patch generation process.

The repair agent, \agentRep, takes the output of both other agents to generate different sets of patches in \autoref{lst:repair-patches}.
All generated patches are validated by both the original test cases to assess their plausibility, and on the newly generated adversarial tests reflecting inferred program intent.   
Our observation is that since the patches are adversarial, it is unlikely that all of them will pass the original test suite. Although incomplete, the developer-provided tests are therefore useful for testing adversarial patch plausibility. In our example, the second patch reflecting adversarial intent  (middle box of \autoref{lst:intent-reasoning}) is correct, while the other two are either overfitting (first box of \autoref{lst:repair-patches}) or non-plausible (third box of \autoref{lst:repair-patches}). 
If only one is a plausible patch, this increases confidence in the inferred intent, and argues for the use of the new tests to supplement the test suite moving forward.
If, on the other hand, all patches are plausible, \approach increases the likelihood that at least one of them will be correct due to adversarial. 


\section{Approach}
\label{sec:approach}

\begin{figure*}[t]
\includegraphics[width=0.97\textwidth]
{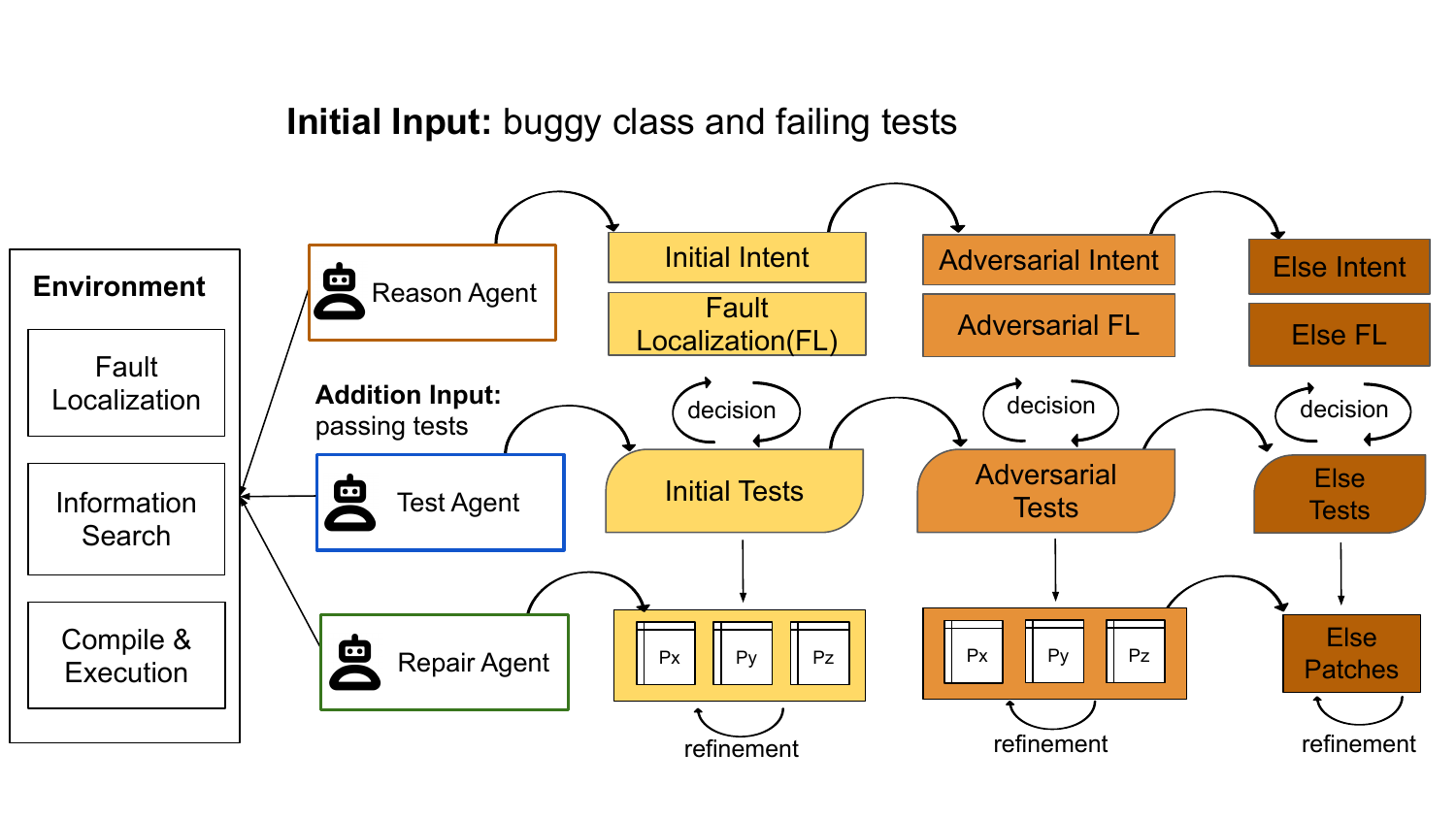} 
\caption{Overview of \approach that shows three agents that are responsible for reasoning -\agentReason, test generation - \agentT, and patch generation - \agentRep. \approach begins with a buggy class and failing tests as inputs, and its outputs include new intents, tests, and candidate patches. }
\label{fig:overview}
\end{figure*}

\autoref{fig:overview} provides an overview of \approach.  
\approach takes as input a program and a set of test cases that identify a bug.  It also is able to interact with an \emph{execution environment} (by, e.g., compiling, executing, or testing the generated code) to collect information as necessitated by the process, or in response to LLM requests. 

\approach then consists of a three-component multi-agent framework: 
\begin{itemize}
\item Reasoning Agent (\agentReason): Reasons about multiple program intents at the function level and locates fault statements that violate such intent. 
\item Test Agent (\agentT): Generates test cases based on the inferred program intent, to verify the adversarial degree of generated intent and act as oracles on the repair process. 
\item Repair Agent (\agentRep): Creates patches based on the located faulty statements and program intent. 
\end{itemize}

A core element of \approach's approach, as illustrated in the motivating example, is the use of \emph{adversarial} reasoning. The key to achieving reasoning is through criticism prompts to tell LLMs the previous answer is insufficient and explicitly to explore diversity, e.g., \emph{The previous answer is not correct, consider alternatives}.
Each agent thus simultaneously explores solutions to the initial repair task, as well as different adversarial formulations of the task.  
This approach is novel compared to related work as follows: 
1) \approach is the first work to reason about multiple, partially adversarial program intents, increasing the likelihood that at least one aligns with the correct intent.
2) \approach is the first work, to our knowledge, to augment test generations during patch generation process, on the opposite of post patch assessments \cite{ICSE18-patchsim, Jooyong-Poracle-TOMSE23} or specific oracle construction (e.g., crash \cite{gao2019crash}.
3) \approach is explicitly guided to explore differences and alternatives,
especially compared with prior conversational approaches \cite{ChatRepair-ISSTA24, ITER}, that limited themselves to iteratively refine generated patches on previous answers.

There are two kinds of \emph{interactions} in \approach: (1) Multi-Agent interactions and (2) Agent-Environment interactions.
In Multi-Agent interactions, \agentReason infers the function intent for \agentT and identifies faulty statements for \agentRep. \agentT generates tests based on the inferred intent to: 1) measure the degree of adversarial behavior between two inferred intents and reject one if the adversarial score falls below a predefined threshold; and 2) use these newly generated adversarial tests to validate the patches produced by \agentRep.
As to Agent-Environment interactions: \agentReason retrieves bug information from the file system to provide additional information requested by the LLMs. \agentT may retrieve sample test cases from the file system, while \agentRep requires the environment to validate the generated patches and collect execution feedback.

The rest of this section describes each agent in \approach and \autoref{tab:prompt} shows prompt examples throughout.

\subsection{\agentReason}

The goal of \agentReason is two-fold: (1) infer program intent to understand expected behavior and (2) locate the buggy statements. 
As shown in \autoref{fig:overview}, \agentReason infers program intents sequentially, first generating an initial intent and then continuously generating adversarial ones based on previous intents, rather than generating them in parallel,  to control each subsequent intent differs from the previous one.

\agentReason takes three inputs:
(a) the source code for the presumed-buggy class;
(b) the failing test cases;
(c) the error messages, which identify the buggy class from the failing test cases where the function is tested but causes test failures.
\agentReason works in two steps. First, it narrows down the buggy class to a finer granularity (functions, constructors, and variables). Second, it infers the intended program behavior and further refines the faulty statements based on this finer code, such as the function level.
As output, \agentReason provides an initial set of program intents and faulty statements that exist in the current version. Each program intent is expected to be explored adversarially and distinctly.

\newcommand{\mmc}[1]{\multicolumn{2}{c|}{\textbf{#1}}}
\newcommand{\mmr}[2]{\multirow{#1}{*}{#2}}

\begin{table}[t!]
    \captionsetup{justification=centering}
    \renewcommand{\arraystretch}{1.2}

    \footnotesize
        \caption{Prompts  used in \approach.}
        \label{tab:prompt}
        \centering
        \begin{tabular}{|p{0.1\textwidth}|p{0.22\textwidth}|p{0.6\textwidth}|}
 	    \hline
      \toprule
 	    \textbf{Agent} &  \textbf{Description} & \textbf{Prompts}  \\ 
      \hline
      \mmr{10}{ \agentReason} &  \mmr{6}{  \makecell{ Fault Localization \\(from class level to \\ finer granularity)}} & \tabitem  P1: Locate the top-3 faulty functions from the given buggy class. \\
      & & \tabitem P2: If the fault does not exist in the function, could you locate the variable definition or other potentially buggy classes? \\
       & &  \tabitem P3: What if the previous answers are incorrect? What alternatives are available? \\
       \cline{2-3}
       & \mmr{4}{  Reason Program Intent} & \tabitem P4: Given the buggy code snippet, can you reason about the program intent and locate the corresponding fault statements? \\
      & &\tabitem  P5: What if the previous intents are incorrect? What alternatives are available? \\
     \hline

 \mmr{12}{ \agentT}  &  \mmr{2}{ Initial Test Generation} & \tabitem P6: Generate N tests based on provided program intent. \\
 \cline{2-3}
 &\mmr{2}{ Criticism Assertions} & \tabitem P7: Are there any assertions that could be wrong? \\
 \cline{2-3}
 & \mmr{4}{ \makecell {Adversarial Test  Generation} } & \tabitem P8: Modify the previous tests based on this intent to maintain the same inputs but generate different outputs. \\

 \cline{2-3}
 &\mmr{2}{Test Prioritization} & \tabitem P9: Rank tests based on confidence of assertion correctness.\\
 \hline
       
    \mmr{10}{ \agentRep}  &   \mmr {6} {Dynamic Precise Prompts} & \tabitem P10: What are the top-3 most likely root causes of this bug-breaking program intent? Answer in X, Y, Z.  \\
    && \tabitem  P11: Generate a patch to repair the bug caused  by X.  \\
     && \tabitem  P12: Generate a patch to repair the bug caused  by Y.  \\
     && \tabitem  P13: Generate a patch to repair the bug caused  by Z.  \\
     \cline{2-3}
     &   \mmr {2} {Patch Refinement} & \tabitem  P14: Refine the patches based on the compilation and execution errors.  \\
     \hline

\end{tabular}
\end{table}

\emph{Adversarial fault localization refined from the class level to statement granularity.}  
\agentReason adversarially explores potential fault localizations by critiquing previous answers and guiding alternative exploration through three adversarial prompts.
As shown in the first row of \autoref{tab:prompt}, \agentReason iteratively narrows the scope of faulty statements from the entire buggy class to finer granularity levels, such as functions (\textbf{P1}). Additionally, \agentReason explicitly guides LLMs to explore statements outside the function level, including constructors and variables, to achieve statement-level fault localization (\textbf{P2}). The LLM is also directed to examine related statements from the initially identified buggy class, as similar issues may exist in other classes, thereby enabling multi-location reasoning (\textbf{P3}). 
Initially, only a single class file is provided. After adversarial prompting and evaluating the error message, the LLM may request additional information. We then supply the requested files to facilitate cross-class fault localization.
%
This design explicitly and systematically explores a diverse set of faulty statements, complementing previous answers. Note that \agentReason is novel in integrating LLM-based fault localization with program repair and reasoning about program behavior. Prior work either assumed perfect fault localization (e.g., RepairAgent~\cite{repairagent}) and others ~\cite{ChatRepair-ISSTA24,alpha-repair-fse22,CURE-icse21,SEQUENCER,CoCoNuT}. Other techniques integrate with spectrum-based fault localization (see, e.g., GenProg \cite {le_goues_genprog_2012} and related techniques ~\cite{RewardRepair-icse22,ye2022selfapr}). This includes agent-like AutoCodeRover \cite{autocoderover}, which considers fault localization and repair as two separate tasks and increases configuration effort.

With the identified faulty functions and statements from the first step, \agentReason begins by generating the initial program intent by asking the LLM, "Can you reason about the program intent and locate the corresponding fault statements?" (\textbf{P4}). Based on the response, \agentReason explicitly guides the LLM to reason about alternative program intents by posing the critical prompt, "What if the previous intents are incorrect? What alternatives are available?" (\textbf{P5}). These critical prompts can be repeated K times until K adversarial intents are generated. 
An example is provided earlier in \autoref{lst:intent-reasoning}, where the response summarizes the program intents in natural language along with the corresponding faulty statements that violate those intents. 
Our insights are twofold. First, \agentReason provides multiple program intents, which are, to some degree, adversarial to each other, thus covering the entire search space of possible intents. This increases the likelihood that at least one of them is correct or close to the original developer's intent. Second, by assuming each individual intent is correct, this approach partially addresses the oracle problem, enabling test oracles to be constructed as long as they satisfy the inferred program intent.


\subsection{\agentT}
\label{sec-app-test}
\agentT takes as input program intents descriptions inferred by \agentReason, and produces as output executable test cases to reflect those behaviors.  
As shown in \autoref{fig:overview},  test generation is a sequential process in which initial tests based on the initial program intent are generated first, followed by tests based on adversarial intents in sequence. Note that each set of test cases is targeted toward validating a particular hypothesis about the function intent. 

\paragraph{There are two objectives in \agentT} 
First, \agentT generates tests to measure the adversarial degree of inferred intents. If the adversarial degree of an intent is low, that intent will be removed, and a new adversarial intent will be generated. 
Second, \agentT generates additional tests to address the overfitting problem by validating the patches generated by \agentRep. Specifically, we validate whether the generated patches are overfitting to a specific intent. When one of the intents is close to the developer's intent, the generated patches are considered correct rather than overfitting.
Note that this use of tests generated as an integral part of repair is, to our knowledge, novel. Overfitting patch assessment tends to be considered a post-processing step in patch generation, or left to manual effort (as in previous agent-based approaches~\cite{repairagent,fix-agent,autocoderover}, which conclude repair attempts upon finding plausible test-passing patches).

\paragraph{Initial test generation and assertion criticism.}
In the initial test generation, \agentT prompts the LLMs to generate N tests based on the initial program intent (\textbf{P6}), establishing a baseline for adversarial tests. However, before modifying these initial tests to create adversarial ones, we use a critique prompt to encourage the LLMs to double-check the correctness of assertions by asking "Are there any assertions that could be wrong?" (\textbf{P7}). To ensure accurate oracles, we remove any tests for which the LLMs are uncertain about their correctness. Since not all LLM-generated test cases are executable. \agentRep attempts to iteratively repair generated tests until they compile, up to three times. Test cases that still do not compile after three repair attempts are discarded.

\paragraph{Adversarial test generation}
Recall that \agentReason produces multiple adversarial program intents. Similarly, \agentT aims to generate \emph{adversarial} tests for these inferred intents. In this context, adversarial tests refer to those with the same input but with expected outputs that vary depending on which intent the program satisfies.
Given the initial set of tests, adversarial prompting directs the LLM to “Modify the expected output according to the adversarial function intent” (\textbf{P8}). This allows us to align all test oracles to the same inputs and count the number of differing outputs. Inspired by mutation testing \cite{jia2010analysis-mutation}, we measure the degree of adversarial conflict in each intent with  $adversarial\_score = \frac{different\_tests}{all\_ tests}$. 
We define an adversarial threshold as $thres= \frac{100\%}{K}$, where 
K is the number of inferred intents. 
For example, in \autoref{fig:overview}, $intents\_1$ and $intents\_2$ produce different outputs in two out of four generated tests, resulting in an \texttt{$adversarial\_score$} of $50\%$.

This design is based on the need to ensure that the test cases are theoretically mutually exclusive, covering the entire intent space, which is our core idea. 
We use the number of adversarial tests to measure the quality of the newly generated inferred intents.
If any inferred intent falls below the threshold, we repeat the generation process until it exceeds the threshold.
Consequently, we increase the likelihood that at least one intent aligns with the correct solution. In this work, we simplify the process by comparing each newly inferred intent with the first generated one.

\paragraph{Test Prioritization.}
We conduct test prioritization for each set of test cases based on the confidence of LLM assertions by asking the LLMs to “rank test cases based on confidence of assertion correctness” (\textbf{P9}). From the ranked list, we select only the top percentage of tests and filter out the rest according to the project's test configuration requirements.

\subsection{\agentRep }

\agentRep takes as input the program intents inferred by \agentReason and the tests generated by \agentT. It aims to generate patches applicable to the input program that fix the bug by ensuring that both the original and adversarial test cases pass. The key difference between \agentRep and prior conversation-based repair approaches is that \agentRep explicitly explores the diversity and differences in the root causes of the bug, whereas prior work either limits themselves in terms of diversity by iterative refining based on a single previous patch \cite{ChatRepair-ISSTA24, ITER} or increases randomness and cost by using random samples in patch generation through adjusting LLM temperatures \cite{hidvégi2024cigar,kong2024contrastrepair}.

\paragraph{Dynamic Adversarial Prompting and Patch Refinement}

\revision{Building upon the previously inferred program intent, our goal is to precisely identify the root causes of bugs to guide repair actions. The purpose is to diversify repair actions: even with the same program intent, multiple different factors may cause a bug. By precisely identifying these root causes, we ensure diversity in repair actions and prevent all generated patches from converging on a single repair strategy.
To this end, \agentRep employs a novel dynamic and precise prompt construction (\textbf{P10}) to explicitly guide patch generation. To identify the root causes of a bug, \agentRep first asks the LLMs to identify the top three most likely issues, denoted as X, Y, and Z. These root causes could include "Null Checks," "Floating Point Precision Issues," "Array Index Errors," "Type Casting," and "Negative Numbers."
These root cause keywords are then passed to the next phase to guide patch generation, ensuring that the process remains explicit and controlled.}


Next, based on these identified root causes, \agentRep sends three separate prompts to guide the LLM in fixing the bug according to each root cause (\textbf{P11}, \textbf{P12}, and \textbf{P13}). Each program's intent leads to three initial patches. However, not all correct patches can be generated in one attempt. Following prior work on conversational LLM approaches \cite{ITER, ChatRepair-ISSTA24}, \agentRep refines previous patches by executing them against the test cases. If errors are found, the error messages are collected and provided to the LLM for further patch refinement (\textbf{P14}). The process is configured to allow a maximum of refinement configuration reaches.



\section{Evaluation}
\label{sec:result}

In this section, we present the experimental setup to evaluate \approach.  Our experiments address the following research questions:

\begin{itemize}
\item \textbf{RQ1} (Effectiveness):  How well does \approach perform overall in fixing bugs, including as compared to prior techniques? 
\item \textbf{RQ2 }(Adversarial Reasoning): To what extent does  adversarial reasoning contribute to \approach's effectiveness in fault localization and patch generation? 
\item \textbf{RQ3 }(Test Cases): To what extent do the test cases generated by \agentT contributes to the quality of the generated patches?
\item \textbf{RQ4 }(Cost): What is the token cost of \approach in generating plausible patches? 
\end{itemize}

We first evaluate the overall effectiveness of \approach in repairing software bugs by comparing its performance with prior related work. Next, we conduct an ablation study to assess the contribution of individual components, such as adversarial reasoning and newly generated test cases, to the overall effectiveness. Finally, we analyze the cost of our approach to provide a comprehensive evaluation of its efficiency.

\begin{table}[t!]
    \captionsetup{justification=centering}
        \renewcommand{\arraystretch}{1.5}

        \caption{ Evaluation benchmarks.}
        \label{tab:benchmark-evaluation}
        \centering
        \scriptsize
        \begin{tabular}{l|lrrrrr}
 	    \hline
 	    \textbf{Datasets} &  \textbf{Projects}  & \textbf{\# Bugs}  \\ 
      \hline
 	        Defects4J 2.0 \cite{Defects4J}  & 17  &  835    \\ 
 	        HumanEval-Java  \cite{jiang2023knod} & 164  &164      \\ 
           \hline
           Total & 181   & 999  \\
 	    \hline
        \end{tabular}
          \end{table}

\subsection{Experimental Setup}
\label{sec:setup}
\approach's core functionality is implemented in Python, interfacing with the GPT-4o API endpoint. We use a default temperature of 1 to promote patch diversity.  We evaluate all generated patches on a 12-core Intel E5-2690V3 CPU at 3.50 GHz with 32GB of RAM, running on Ubuntu 20.04.3 LTS and utilizing OpenJDK Java 64-Bit Server version 1.8.0\_312. 

In the experiment configuration, we set the number of inferred intents to 
$K=3$. We aim to locate the top three faulty functions in the provided buggy class and two alternative answers in the constructions, definitions, and other classes, for a total of five. For test generation, we take 70\% of the ranked adversarial tests based on the confidence of assertions and filter out the remaining ones. For patch generation, we identify the top three causes that are inconsistent with the inferred program intent, and we allow each initial patch three rounds of refinement.

\subsubsection{Benchmarks}

We evaluate \approach on the widely used benchmarks Defects4J 2.0~\cite{Defects4J} and HumanEval-Java~\cite{jiang2023impact}. 
\autoref{tab:benchmark-evaluation}
 summarizes details of the considered benchmarks.
Defects4J 2.0 consists of 835 bugs from 17 real-world projects and has been widely used in related work, enabling us to conduct a fair comparison. Additionally, we evaluate HumanEval-Java. 
In HumanEval-Java, developers converted Python programs and their associated test cases from HumanEval into Java programs and JUnit test cases, as well as some bugs were intentionally introduced into the correct Java programs.
\revision{
HumanEval-Java was released after the data collection used for training GPT-3.5, providing a new benchmark for evaluating the model's ability to handle Java coding tasks.
}

\subsubsection{ Patch Quality Evaluation}

We compare \approach against the state-of-the-art program repair baselines on open-source projects. 
We measure \approach and baseline effectiveness at bug repair according to standard metrics in the APR literature:

\begin{itemize}
\item Plausible: the number of bugs that include at least one patch that makes the original developer-written test cases pass.
\item  Correct: the number of bugs that include at least one correct patch, as discussed below.  
\item  \revision{$Top@N$: A metric that evaluates whether at least one of the top-$n$ generated patches is correct. A patch is considered correct if it passes both the original test cases and the automatically generated test cases. }
\end{itemize}

We measure APR effectiveness in terms of the number of both plausible (test-passing) patches and correct patches.  Patches that overfit to the tests but fail to generalize to the desired specification are a well-known problem in automatic program repair~\cite{cure-worse-15, kali-issta15}.  
We take multiple approaches to evaluate patch correctness. First, we assess whether the patch is plausible that makes the original test cases pass \cite{kali-issta15}. If a patch is plausible, we further identify whether it is an exact match to the developer-written ground truth patch with string comparison, ignoring whitespace differences.

For the remaining plausible patches that do not exactly match the ground truth, we employ an LLM to assess their correctness based on whether they pass LLM-generated tests for based on one specific program intent. If a patch passes its intended test case, it is considered a likely-correct patch; otherwise, it is considered likely-overfitting.
Finally, we manually assess patch correctness following prior work~\cite{ChatRepair-ISSTA24,acs,genpro2009,CURE-icse21}. Two authors independently cross-check the correctness of these patches~\cite{CURE-icse21,ye2022selfapr,jiang2023knod,repairagent}. If both authors agree that a patch is semantically equivalent to the ground truth, it is considered believed-correct. These, along with the exact match patches, form the set of correct patches.

\subsubsection{Methodology}
We discuss our methodology for each research question below.

\paragraph{Overall effectiveness (RQ1)}
We compare \approach with related work across different repair families, including fine-tuning-based methods: SelfAPR~\cite{ye2022selfapr}, AlphaRepair~\cite{alpha-repair-fse22}, ITER \cite{ITER}; conversation-based methods: ChatRepair \cite{ChatRepair-ISSTA24}, Cigar \cite{hidvégi2024cigar}, ContrastRepair \cite{kong2024contrastrepair}, and a baseline from GPT-4; as well as the agent-based approach: RepairAgent \cite{repairagent}. We reimplemented experiments for ChatRepair \cite{ChatRepair-ISSTA24} based on GPT-4 and also conducted a baseline experiment using GPT-4 without execution feedback. For the remaining approaches, we report quantitative results from the corresponding papers and repositories. Plausible and correct repair results are reported as discussed above.
In the experiment assuming perfect fault localization, we follow prior work \cite{repairagent,kong2024contrastrepair,ChatRepair-ISSTA24} by providing the buggy statements and their contextual code to the LLMs. For a more realistic comparison, we provide a complete buggy class tested by the failing test cases, allowing the LLM to achieve finer granularity at both the function and statement levels.
Existing related work evaluated on HumanEval-Java is mostly based on LLM approaches. 
\revision{This is because HumanEval-Java was released recently, after the data collection used for training GPT-3.5, meaning that earlier approaches were not able to evaluate it at the time they published their paper.}

\paragraph{Adversarial Reasoning (RQ2)}
We study the effectiveness of adversarial reasoning by: 
1) Measuring the adversarial degree between inferred intents to assess how different and distinct they are. This is based on the number of distinct test oracles generated from the same inputs. The evaluation is performed during the initial intent generation, without regenerating intents if the score falls below the threshold. 
2) Comparing the alignment between inferred intents and ground-truth intents extracted from the human-written patch. We first use an LLM to generate the ground-truth intent from the patch, then assess semantic equivalence using the GPT-4o.
3) Evaluating its effectiveness as a fault localization and patch generation technique. For buggy statements spanning multiple locations, identifying any part of the buggy statement is considered correct, as it provides a strong starting point for patch generation. This is justified by prior work \cite{ITER}, which shows that subsequent buggy statements can be discovered iteratively. For patch generation, we compare the number of successful patches generated from the initial intent and the two adversarial intents to evaluate how adversarial reasoning promotes diversity. All three analyses involve substantial manual effort. To make this feasible, we conduct the study on 300 randomly selected bugs from Defects4J 2.0.

\paragraph{Test Generation (RQ3)}
In this research question, we explore to what extent newly generated tests can successfully identify and filter out overfitting patches. To evaluate this, we adopt a two-step approach that combines LLM-based selection with manual confirmation. Specifically, we employ GPT-4o, which takes as input a set of plausible patches and a set of newly generated tests. First, we perform a sanity check on the generated tests by assessing their correctness and removing those with low-confidence oracles. Next, we rank the tests and select the top 70\% from the ranked list. These selected tests are then executed against the plausible patches to identify likely overfitting patches, defined as those that fail the tests. For all patches flagged as likely overfitting by the LLM, we conduct a final manual analysis to determine the actual number of true positives.

\paragraph{Cost (RQ4)}
In RQ4, we conduct a detailed analysis of the cost associated with \approach by comparing it to several related methods, including ChatRepair\cite{ChatRepair-ISSTA24}, Cigar \cite{hidvégi2024cigar}, and RepairAgent \cite{repairagent}, all of which rely heavily on large language model (LLM) APIs for automated program repair tasks. Our comparison focuses specifically on the token cost, which refers to the average number of tokens consumed by each approach to repair a  bug in the Defect4J benchmark. We evaluate the token cost in real-world software projects, thus we do not evaluate the token cost on HumanEval-Java. By evaluating token usage in this context, we aim to provide a clear understanding of the computational resources required by each method and to highlight the efficiency and practicality of \approach in real-world software engineering scenarios.

\begin{table*}[t]
\footnotesize

\caption{Comparison with state-of-the-art on three evaluation benchmarks: Defect4J 2.0 and HumanEval-Java. A ‘*’ indicates our reimplementation results, and a ‘–’ indicates that the result is not available in the literature. The best performance is shown in bold.}
\renewcommand{\arraystretch}{1.}
    \label{tab:rq1-comparison-claire}
    \begin{tabular}{ll|rr|rrr}
\toprule
\mmc{\multirow{2}{*}{\textbf{Approach}}} &\multicolumn{2}{c}{\textbf{Perfect FL}}& & \textbf{Realistic FL}  \\
       &   & \textbf{plausible}  & \textbf{correct}  & \textbf{plausible}  & \textbf{correct}   \\
\midrule
& \multicolumn{4}{c}{Defects4J 2.0 (835 bugs)} \\
\midrule
\mmr{2}{Fine-tuning-based} & SelfAPR    \cite{ye2022selfapr}     & 130 & 110 & 107 & 67  \\
& AlphaRepair  \cite{alpha-repair-fse22}& 109 & 79  & 90 & 50  \\
& ITER \cite{ITER} & - &  -  &119 & 74 \\

\mmr{3}{LLM Conversation-based} & ChatRepair  \cite{ChatRepair-ISSTA24} &\cellcolor{white} -  & \cellcolor{white}162 
 &\cellcolor{white} -&\cellcolor{white} -  \\
  &\cellcolor{white}Cigar \cite{hidvégi2024cigar} &\cellcolor{white}- &\cellcolor{white}- & \cellcolor{white}\textbf{185} & \cellcolor{white}69   \\   
  &\cellcolor{white}ContrastRepair \cite{kong2024contrastrepair}  &  \cellcolor{white} \textbf{201} & \cellcolor{white}143 &\cellcolor{white}-&\cellcolor{white}- &  \\

 \mmr{2}{Agent-based}& RepairAgent \cite{repairagent} & 186 & \textbf{164} &- & - \\  
& \textbf{\approach} & 180 &  141 &135 & \textbf{77}  \\
 \midrule

& \multicolumn{4}{c}{HumanEval-Java (164 bugs)} \\
\midrule
\multirow{5}{*}{\textbf{LLM Conversation-based}}  & GPT-4  *   &  136 &  127 & 124 & 87  \\
 & ChatRepair *  \cite{ChatRepair-ISSTA24}  &137 & 130  & 126 &  88  \\

 & ContrastRepair \cite{kong2024contrastrepair} &151 & 137 & - & - \\

  & Cigar \cite{hidvégi2024cigar} & - & - & \textbf{152} & 102\\
  & \textbf{\approach} & \textbf{154 }& \textbf{140 } &  146  & \textbf{ 105 }  \\

\bottomrule        

    \end{tabular}
\end{table*}

\subsection{RQ1: \textit{\approach} Effectiveness}


\autoref{tab:rq1-comparison-claire} compares the effectiveness of \approach with state-of-the-art in two considered benchmark Defects4J 2.0 and HumanEval-Java. 
The second column lists the existing APR approaches from three categories: fine-tuning-based, LLM-based and Agent-based approaches. Their effectiveness results respectively based on perfect fault localization and realistic fault localization are given in the third and fourth column groups. The best performance number is highlighted in bold in the table. 
For example, in the last row of the Defects4J 2.0 benchmark, \approach generated plausible patches for 135  bugs and correctly repaired 77 of them with realistic fault localization.

\approach achieves the highest number of correct repairs in both the Defects4J 2.0 benchmark (77 of 835 bugs repaired) and the HumanEval-Java benchmark (105 of 164 bugs repaired) under the realistic fault localization configuration. 
For Defects4J 2.0 benchmark, these numbers increase to 180 and 141, respectively, when assuming perfect fault localization is known. For HumanEval-Java benchmark, these numbers increase to 154 and 140, respectively, when assuming perfect fault localization is known.
Our result shows the overall effectiveness of \approach.

While \approach does not produce the highest number of plausible patches, this is due to its stronger validation criteria: patches are assessed against both the original and adversarial test cases, where adversarial tests effectively filter out overfitting patches. In contrast, prior work only validates against original test cases, resulting in a higher number of overfitting patches that cannot be eliminated.



We present a case study to demonstrate the effectiveness of \approach.
\autoref{lst:rq1-case} shows patches generated based on the same program intent for the \texttt{add\_elements} program in HumanEval-Java.
\autoref{rq1-example-precise-bug-descrip} provides an initial inferred program intent and identifies the faulty statement in which the given code checks for at least two digits, which violates the inferred intent  of at most two digits (see the comment in \autoref{rq1-example-precise-bug-descrip}).

\autoref{lst:rq1-case-init-patch} shows the initial patch generated by \agentRep. Although this patch is intended to fix the issue, it still contains two bugs. First, the patch incorrectly identifies all numbers under \textit{100} as having two digits, failing to account for negative numbers that may have more than two digits. Second, the initial intent and its patch fail to correct the buggy code that subtracts from, rather than adds to, the final result.

The adversarial intent, shown in \autoref{rq1-example-adver-intent}, critiques the initial intent and patch and infers a new intent that differs from the initial intent. This leads to the generation of an adversarial patch, shown in \autoref{lst:rq1-case-refine-patch}. Based on the inferred program intent, \approach also generates an additional test case \autoref{rq1-example-test} to reveal the behavioral difference between the initial patch and the adversarial patch: the expected output is '-4' (making the adversarial patch pass), while the initial patch's actual output is '-16' (which fails this test case).


We can see \approach provides rich information, including inferred program intent and newly generated test cases, which offer valuable resources for developers to understand the patches. This is a significant distinction from prior work, which typically provides only patches.
To our knowledge, \approach is the only framework that integrates inferred program intents, tests, and patches. This paradigm shift enhances how developers engage with APR patches, enabling them from understanding the patch and choosing one program intent that most closely aligns with their original intent.

\begin{figure}[t!]

\noindent\begin{minipage}[b]{.95\textwidth}

\begin{tcolorbox}
[colback=white,colframe=black,arc=0pt,boxrule=0.5pt,title=  An inferred program intent and located faulty statements:,boxsep=2pt,left=1pt,right=1pt,top=1pt,bottom=1pt,fonttitle=\footnotesize]
\footnotesize
\textbf{Program Intent:} Given an array of integers arr and an integer k, the program should return the sum of elements that have at most two digits, within the first k elements of arr. For this calculation, any negative sign should count as a digit, so negative numbers up to -99 (e.g., -5, -23) qualify as "two-digit numbers" under this rule.

\textbf{Suspicious Faulty Statements}

\colorbox{red!30!}{ -   if( Integer.toString (arr.get(i)).length() >= 2) \{}  \textcolor{deepgreen}{//bug: at least instead of at most }

\colorbox{red!00!}{   -    result -= arr.get(i);} 

\colorbox{red!00!}{   -    \} }
\end{tcolorbox}
\subcaption{\scriptsize An inferred program intent and faulty statements}
\label{rq1-example-precise-bug-descrip}    \end{minipage}
    
\begin{minipage}[t]{.95\textwidth}
\begin{minipage}[t]{.49\textwidth}
\begin{tcolorbox}
[colback=white,colframe=black,arc=0pt,boxrule=0.5pt,title=An initial patch:,boxsep=2pt,left=1pt,right=1pt,top=1pt,bottom=1pt,fonttitle=\footnotesize]
\scriptsize

\textcolor{deepgreen}{//Bug: -10 will be incorrectly count}
\colorbox{red!30!}{ -   if( Integer.toString (arr.get(i)).length() >= 2) \{}  
 \colorbox{green!30!}{ + int num = arr.get(i); } 

\colorbox{green!30!}{+ if (Math.abs(num) < 100) \{ }
\end{tcolorbox}
\subcaption{\scriptsize An initial patch  }
\label{lst:rq1-case-init-patch}            
\end{minipage}
\begin{minipage}[t]{.5\textwidth}
\begin{tcolorbox}
[colback=white,colframe=black,arc=0pt,boxrule=0.5pt,title=An adversarial inferred program intent:,boxsep=2pt,left=1pt,right=1pt,top=1pt,bottom=1pt,fonttitle=\footnotesize]
\scriptsize

\textcolor{deepgreen}{//bug 1: at least instead of at most}

\colorbox{red!30!}{ 
-   if( Integer.toString (arr.get(i)).length() >= 2) \{}  \textcolor{deepgreen}{ }

\textcolor{deepgreen}{//bug 2: Subtraction instead of addition }

\colorbox{red!30!}{   -    result -= arr.get(i);} 

\end{tcolorbox}
\subcaption{\scriptsize An adversarial inferred program intent and faulty statements}
\label{rq1-example-adver-intent} 

\end{minipage}
\end{minipage}

\begin{minipage}[t]{.95\textwidth}
\begin{minipage}[t]{.49\textwidth}
\begin{tcolorbox}
[colback=white,colframe=black,arc=0pt,boxrule=0.5pt,title=An adversarial patch:,boxsep=2pt,left=1pt,right=1pt,top=1pt,bottom=1pt,fonttitle=\footnotesize]
\scriptsize
\colorbox{green!30!}{ + int num = arr.get(i);}

 \colorbox{green!30!}{+  String numStr = Integer.toString(num); }
 \colorbox{green!30!}{+   int digitCount = numStr.length(); }
 
\textcolor{deepgreen}{//Check at most two digits}

\colorbox{green!30!}{+  if (digitCount <= 2) \{ }

\colorbox{green!30!}{+       result += num;  }\textcolor{deepgreen}{//Addition instead of subtraction} 

\colorbox{green!30!}{+   \} }

\end{tcolorbox}
\subcaption{ \scriptsize An adversarial patch  }
\label{lst:rq1-case-refine-patch}              
\end{minipage}
\begin{minipage}[t]{.5\textwidth}
\begin{tcolorbox}
[colback=white,colframe=black,arc=0pt,boxrule=0.5pt,title=A generated test case:,boxsep=2pt,left=1pt,right=1pt,top=1pt,bottom=1pt,fonttitle=\footnotesize]
\scriptsize

@Test

public void testAddElements() \{

 ArrayList arr = new ArrayList(Arrays.
 asList(-1, -12, -3));
 
 int k = 3;
 
 \textcolor{deepgreen}{// Expected output: -1 + (-3)}
 
 assertEquals(-4, add\_elements(arr, k));   
 
\footnotesize \}

\end{tcolorbox}

\subcaption{\scriptsize A test case that makes (b) initial patch fail and (d) adversarial patch pass}
\label{rq1-example-test} 

\end{minipage}
\end{minipage}



\hfill
\hfill
\captionsetup{justification=centering}
\caption{ A correct patch generated by \approach.}
\label{lst:rq1-case} 

\end{figure}




\begin{tcolorbox}
[colback=white,colframe=black,arc=0pt,boxrule=0.5pt,title=Answer to RQ1:,boxsep=2pt,left=1pt,right=1pt,top=1pt,bottom=1pt,fonttitle=\bfseries]
 
\approach repairs 77 in Defects4J and 105 in HumanEval-Java benchmarks. This success is due to effective adversarial reasoning on program intents and dynamic, root-cause-specific prompt construction. Additionally, \approach offers comprehensive resources for function reasoning, bug analysis, and test cases, providing a unique package in program repair approaches. This design shifts the paradigm by enabling developers to understand APR patches and select program intents more effectively.
\end{tcolorbox}

\subsection{RQ2: Adversarial Intent Effectiveness}

\begin{table}
 \renewcommand{\arraystretch}{1.2}   \captionsetup{justification=centering}
\caption{Program intent alignment statistics.}
\footnotesize
\label{tab:intent_stats}
\begin{tabular}{lrr}
\toprule
                                Scenario &  Count &  Percentage  \\
\midrule
        All three intents are aligned &    75 &            25.0 \% \\
     Only the first intent is aligned &     38 &            12.7  \% \\
    Only the second intent is aligned &     21 &             7.0  \% \\
     Only the third intent is aligned &     21 &             7.0  \% \\
Both the first and second are aligned &     56 &            18.6   \% \\
 Both the first and third are aligned &     17 &             5.7  \% \\
Both the second and third are aligned &     17 &             5.7  \% \\
  \hline
  At least one intent is considered aligned (total) &    245 &            81.7  \% \\
  None of the three intents is considered aligned   & 55 &  18.3  \%\\
\bottomrule
\end{tabular}
\end{table}

\begin{figure}[t!]
\includegraphics[width=0.6\textwidth]{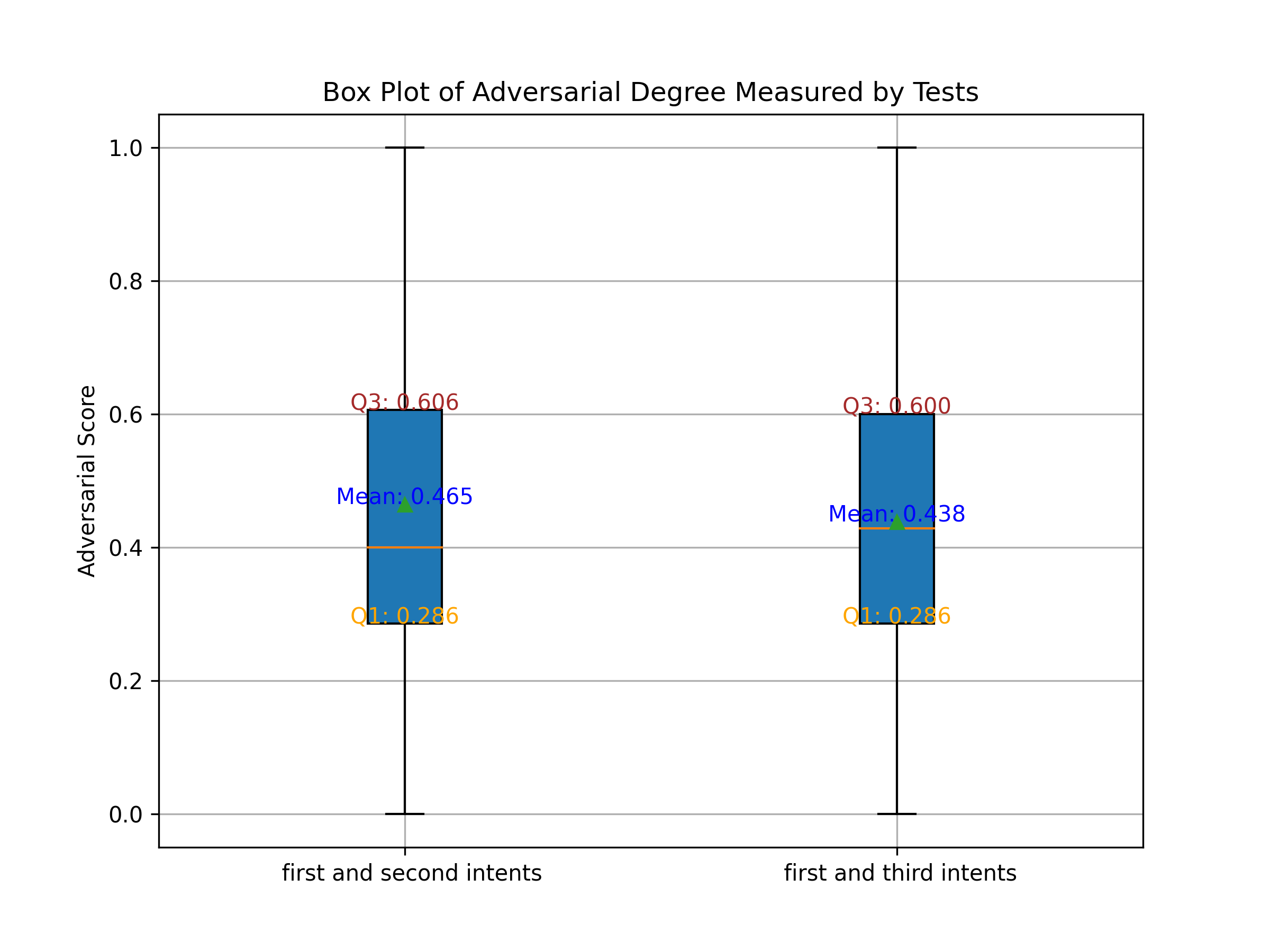} 
\centering
 \captionsetup{justification=centering}
 \caption{ Adversarial degree between the inferred intents.}
\label{fig:ad-intent}
\end{figure}

We analyze the effectiveness of adversarial intents based on 300 randomly selected bugs from the Defects4J benchmark.
\revision{First, we quantify the adversarial score between different inferred intents. As shown in \autoref{fig:ad-intent}, the majority of adversarial scores fall within the range of 28.6\% to 60.0\%. The average adversarial score ranges from 43.8\% to 46.5\%. This data confirms the adversarial nature between different intents.
}

\revision{Second, we measure the extent to which the generated intents align with the expected program intent (ground truth intent). 
\autoref{tab:intent_stats} presents the analysis, showing how well the inferred intents match the expected ground truth  intent.
As shown in the two last rows of \autoref{tab:intent_stats}, our analysis indicates that 81.7\% of cases with at least one inferred intent are aligned with the ground-truth intent, while only 18.3\% of cases have no correctly inferred program intents.

Specifically, it highlights the significant benefits of using multiple intents. If only a single inferred intent is used, the best single-intent match (the first intent) achieves an accuracy of 62.0\% (the sum of 25.0\%, 12.7\%, 18.6\%, and 5.7\%). In contrast, when leveraging multiple intents, at least one correct match is found in 81.7\% of cases. The second and third inferred intents complement the first intent, accounting for the remaining 19.7\% of  matches.
}


\autoref{sec-rq2-ar} presents an ablation study on the effectiveness of \approach with and without adversarial intents in fault localization and patch generation. The last row provides a summary of the study. We can see that with adversarial reasoning, fault localization precision increases by 13.8\%, and patch generation improves by 19.4\%. 
This confirms the effectiveness of reasoning adversarial intents for both fault localization and patch generation. We make the following implications.

Fault localization performance improves as adversarial intents encourage exploration at both function and statement levels, extending beyond the provided class.
First, adversarial reasoning examines a broader and more varied search space for faulty code, not merely by generating more candidate faulty statements. Our experiment shows that simply increasing the number of candidates does not necessarily improve the final result.
Second, adversarial reasoning guides the LLM to locate buggy statements outside functions, such as variable definitions, constructors, and other classes.

\begin{table*}[t!]
    \captionsetup{justification=centering}
            \renewcommand{\arraystretch}{1.1}
\footnotesize
        \caption{Adversarial intent  effectiveness. }
        \label{tab:benchmark}
        \begin{tabular}{l|rr|rrrrrr}

 	    \hline

\multirow{3}{*}{Projects} & \multicolumn{2}{c}{\textbf{Fault Localization} } & & \multicolumn{2}{c}{\textbf{Patch Generation}} \\

      & Initial Intent &  + Adversarial Intents  & &
   Initial Intent & + Adversarial Intents \\
 

      \hline


\hline
Total & 64  &  75 (13.8\%) & & 36 & 43 (+19.4\%)\\
          
\hline
\end{tabular}
\label{sec-rq2-ar}
\end{table*}

Patch generation performance improves as adversarial intents increase the likelihood of the search space containing a correct patch. 
We observe that successful patches benefit from prompts that are both adversarial and precise. The number of correctly repaired bugs increased from 36 to 43 by incorporating adversarial intents.  Adversarial intents explore a diverse range of expected program behaviors, avoiding repeated flawed attempts and effectively pinpointing root causes with accuracy.

\begin{tcolorbox}
[colback=white,colframe=black,arc=0pt,boxrule=0.5pt,title=Answer to RQ2:,boxsep=2pt,left=1pt,right=1pt,top=1pt,bottom=1pt,fonttitle=\bfseries]

Adversarial intent reasoning significantly boosts fault localization precision by 13.8\% and enhances patch generation by 19.4\%.
Adversarial reasoning expands the search space by encouraging exploration beyond function-level constraints, effectively locating buggy statements and enhancing patch accuracy. Additionally, the agent-driven loop facilitates broader exploration by providing context, helping avoid dead-end responses.

\end{tcolorbox}

\subsection{RQ3: Tests and Overfitting}

\autoref{tab:rq3} shows the number of tests generated by \agentT and the number of bugs with overfitting patches discarded by adversarial tests, where overfitting is associated with the original tests. The second column shows the initial number of generated tests, while the third column indicates the number of low-confidence assertions, as identified by evaluating the correctness of assertions in generated patches, and we prioritize tests based on the LLM's assertion confidence. The number of bugs with likely-overfitting patches that are discarded are given in the last column.

We make the following observations from \autoref{tab:rq3}. 
First, \approach discards likely-overfitting patches for 12 bugs in Defects4J 2.0 and 7 bugs in HumanEval-Java, confirming the effectiveness of adversarial testing in alleviating overfitting.
Second, using criticism prompts to ask LLMs to double-check assertions and identify potentially incorrect ones is effective, helping us filter out 23.8\% to 28.4\% of low-confidence tests.

\begin{tcolorbox}
[colback=white,colframe=black,arc=0pt,boxrule=0.5pt,title=Answer to RQ3:,boxsep=2pt,left=1pt,right=1pt,top=1pt,bottom=1pt,fonttitle=\bfseries]
AdverIntent-Agent effectively discards overfitting patches for 12 bugs in Defects4J 2.0 and 7 bugs in HumanEval-Java. Test generation faces a low compilation rate, with variation between complex and simpler programs, largely due to missing dependencies. 
 
\end{tcolorbox}

\subsection{RQ4: Cost of \approach}

\autoref{table-time-cost} compares the average token cost across four approaches: Cigar, RepairAgent, ChatRepair, and \approach. On average, \approach uses 438K tokens per bug. Its token usage is higher than both RepairAgent and Cigar but remains lower than ChatRepair. 

The main reason for the increased cost compared to RepairAgent is that \approach performs realistic fault localization and automated test generation for each bug, with test generation representing the primary contributor to the overall token cost, a step that is not included in RepairAgent’s workflow. 

Compared to Cigar, \approach’s higher token cost results from its use of advanced prompts for reasoning about program intent and generating tests, both of which are key innovations in our approach. However, if we consider only the patch generation phase, \approach is actually more cost-effective than Cigar because it requests three patches per program intent and root cause, whereas Cigar samples 50 candidate patches per iteration, leading to a much greater token expenditure for Cigar in this stage. 

Despite these additional costs, \approach remains more efficient than ChatRepair. ChatRepair’s iterative approach incorporates all historical information into each prompt, causing token consumption to increase rapidly as iterations accumulate, sometimes reaching into the hundreds. By contrast, \approach is designed to limit itself to three rounds of test and patch refinements, which manages overall token usage

\approach achieves a balance between realism and efficiency, using more tokens for enhanced reasoning and test generation than some baselines, but remaining substantially more cost-effective than highly iterative methods like ChatRepair.

\begin{table*}[t!]
\footnotesize
    \captionsetup{justification=centering}
            \renewcommand{\arraystretch}{1.2}

        \caption{Generated test cases and overfitting detection.}
        \label{tab:rq3}
        \centering
        \begin{tabular}{l|rrr|rl}

\hline

& \multicolumn{3}{c}{Test Cases}  &Likely-overfitting  \\


&Initial     & Low-Confidence   &  Rank Threshold    &     \\
\hline
Defects4J 2.0 & 25050   &  28.4\% & 70\% & 12  \\
HumanEval-Java& 4920  & 23.8\% & 70\%   & 7  \\

\hline
\end{tabular}
\end{table*}

\begin{tcolorbox}
[colback=white,colframe=black,arc=0pt,boxrule=0.5pt,title=Answer to RQ4:,boxsep=2pt,left=1pt,right=1pt,top=1pt,bottom=1pt,fonttitle=\bfseries]
\approach’s cost increases due to program intent reasoning and test generation, yet remains manageable because of the fewer number of iterations. 
 
\end{tcolorbox}

\section{Related Work}
\label{sec:related}

\subsection{Automated Program Repair}
\emph{Code Transformation.} Source code transformation to alter program behavior has been the main research focus of many search-based \cite{sofix,genpro2009,le_goues_genprog_2012,Yuan2017ARJAAR,Arja-e-Tosem20,ssFix,long_staged_2015,tbar,acs,martinez2016astor,saha2019harnessing}, constraint-based \cite{semfix,Angelixicse16,nopol,concolic-repair-PLDI21,relifix-shinwei-icse15,directfix}, and learning-based repair \cite{tse24-Search-Generate-Modify,codit-tse20,CoCoNuT,DLFix,SEQUENCER,CURE-icse21,jiang2023knod,RewardRepair-icse22} techniques for a decade. The emphasis lies in constructing syntactically and semantically correct code transformations, taking into account their expressiveness and effectiveness.

\emph{ Rich Information Beyond Code.} Recently, more work considered information beyond code transformation, by including 
commit message \cite{modit-ase21}, compiler and test feedback \cite{ye2022selfapr,kong2024contrastrepair},
domain knowledge \cite{tosem24-domain,ye2022selfapr} and iterative patch refinement \cite{ITER}. 
Chen et al. \cite{chen2024teaching}
improves LMs code generation accuracy by injecting feedback messages generated by the LM itself based on its code execution results.

\emph{Conversational-based Repairs.} Recent advancements in Large Language Models (LLMs) have paved the way for innovative approaches to automated program repair (APR). ChatRepair\cite{ChatRepair-ISSTA24} pioneered a conversational repair method based on LLMs, iteratively generating patches by collecting bug context information and patch execution results. Cigar\cite{hidvégi2024cigar} further optimized the repair process by designing prompts and reboot strategies to generate diverse patches while reducing token consumption. 
ContrastRepair \cite{kong2024contrastrepair} takes an LLM-based iterative approach for APR. Their prompt includes a failing test and a similar passing test, where a passing test is either selected from the existing suite or generated using type-aware mutation.
CHATDBG \cite{levin2024chatdbg} shifted the focus from fully automated repair to a co-pilot for debugging, integrating LLMs with standard debuggers to enhance their capabilities.

\emph{Patch Correctness Assessment}. Existing methods for assessing overfitting patches typically classify these patches separately and address them in post-processing. Prior work mainly focuses on static code analysis \cite{ye2021ods, TianASE20, ASE20Wang, anti-pattern} and test generation and execution analysis \cite{gao2019crash, tian2021failtest, ICSE18-patchsim, issta17-difftgen}.


Our work falls into the category of conversational-based repairs. Unlike previous prompting-based repair approaches, ours is the first to address the overfitting patch problem during the patch generation phase rather than as a post-generation step.

\subsection{LLM-Based Agent for Code}
There is a set of parallel LLM-based agents designed for repairing software issues related to our research \cite{yang2024sweagent, zhang2024codeagent, autocoderover}. Yang et al. \cite{yang2024sweagent} propose SWE-Agent for repairing GitHub issues. AutoCodeRover \cite{autocoderover} combines LLMs with advanced code search techniques to address GitHub issues through program modification or patch generation. Zhang et al. \cite{zhang2024codeagent} introduce CodeAgent, a novel LLM-based agent framework that employs external tools for repository-level code generation, integrating five programming tools for information retrieval, code symbol navigation, and code testing. Bouzenia et al. \cite{repairagent} propose RepairAgent, which treats the LLM as an agent capable of autonomously planning and executing actions to fix bugs by invoking suitable tools. Our work similarly uses a decision-action-planning loop with multiple iterations and dynamic prompts based on command outputs, with the novel component of reasoning about multiple adversarial intents to explore diverse solutions.

\begin{table}[t!]
\renewcommand{\arraystretch}{1.5}
\footnotesize
\centering
    \captionsetup{justification=centering}

\caption{Token cost analysis of \approach.}

\begin{tabular}{cccccccc}

\hline

 & Cigar & RepairAgent  &  ChatRepair & \approach  \\
\hline
Tokens   & 127K & 270K & 467K  & 438K  \\
\hline
\end{tabular}

\label{table-time-cost}
\end{table}



\subsection{Program Reasoning}

Program intent reasoning is a critical area of research within software engineering, encompassing various techniques and methodologies aimed at understanding, analyzing, and improving software programs. One notable approach is symbolic execution \cite{varfix,sergey-test-equivalence-tosem18}, which systematically explores program paths to identify potential errors or vulnerabilities.
Formal methods provide mathematical techniques for reasoning about program correctness and behavior \cite{SMT}.
\revision{SpecRover by Ruan et al. \cite{SpecRover-icse25} is a closely related work that extracts code intents via LLMs. This work shares a similar idea with ours by first extracting program intent and then using it to guide patch generation. However, our approach conceptually generates multiple diverse program intents adversarially, maximizing the likelihood that at least one inferred program is correct, whereas SpecRover infers only the most likely program intent.}

In the general setting of machine learning for program reasoning, several works propose  Chain-of-Thought  \cite{yang2023chainofthought}, Tree-of-Thought \cite{yao2023treeofthoughts} and Graph-of-Thought \cite{Graph-of-Thoughts} to
logical reasoning code generation tasks by breaking them down into understandable intermediate steps, enabling LLMs to handle each step individually.  
Unlike prior work, our approach reasons about program intent at the function specification level and incorporates multiple adversarial intents to ensure diverse solutions are explored, which is novel to our knowledge.

\section{Threats to Validity}
\label{sec:threats}

We acknowledge several potential threats to validity:
1) \texttt{Manual patch validation}.
The first internal threat arises from the manual validation of believed-correct patches against the reference developer patch.
To address this, we carefully examined and discussed each patch, and we make the patches publicly available.
2) \texttt{Data leakage}. Using GPT-4o for \approach raises data leakage concerns due to its undisclosed training data; however, evaluation on the HumanEval-Java dataset helps strengthen internal validity.
3) \texttt{Non-deterministic of LLMs.}
Non-deterministic language model outcomes pose a threat. We mitigate this by evaluating a substantial number of bugs and providing interaction logs for reproducibility and transparency.
4) \texttt{Limited programming languages.}
Our experiments are limited to Java projects, which threatens external validity, as results may not generalize to other programming languages or projects.
To mitigate this, we include two benchmarks, Defects4J 2.0 and HumanEval-Java, covering 181 diverse projects with various bug types, lengths, and complexities, which strengthens the validity of our findings.
5) \texttt{Hallucinations and mistakes of LLMs.} We use LLMs for test generation based on specific program intent, but the generated oracles can sometimes be incorrect and difficult to detect. LLMs are also used to identify overfitting patches, which may result in further mistakes. We mitigate this risk by manually reviewing the results of LLM-based detection.

\section{Conclusion}
\label{sec:conclusion}

We  present \approach, a multi-agent system that advances automated program repair by focusing on inferred program intent through adversarial reasoning. Unlike traditional APR approaches, AdverIntent-Agent generates diverse program intents, adversarial tests, and patches guided by precise root causes to reduce overfitting and improve alignment with developer intent. Evaluations on Defects4J 2.0 and HumanEval-Java show that AdverIntent-Agent repairs more bugs than previous methods, underscoring its potential to enhance APR with intent-driven, developer-oriented solutions. 
Our work suggests a new paradigm in patch assessment that considers developer acceptance of APR patches, shifting to asking developers to analyze the patches and select the inferred program intent that best aligns with their original intent.

\begin{acks}
This work was partially supported by The Wallenberg Foundation and KAW Postdoctoral Scholarship Program - KAW 2022.0368. The computations and data handling were enabled by the supercomputing resource Berzelius provided by National Supercomputer Centre at Linköping University and the KAW foundation: 2024 NAISS GPU Grant Berzelius-2024-131.
\end{acks}

 \bibliographystyle{ACM-Reference-Format}
 \bibliography{reference.bib}

\end{document}